\documentstyle[aps,amssymb,epsfig,twocolumn]{revtex}  
\begin{document}
\textheight 230mm
\wideabs
{
\title{\LARGE 
{\sf Evolution of topological order in Xe films on a quasicrystal surface}}
\author{
Stefano Curtarolo$^{1,3}$, Wahyu Setyawan$^{1}$, Nicola Ferralis$^{2}$, Renee D. Diehl$^{2}$, Milton W. Cole$^{2}$}
\address{
$^1$Department of Mechanical Engineering and Materials Science, Duke University, Durham, NC 27708 \\
$^2$Department of Physics and Materials Research Institute, Penn State University, University Park, PA 16801\\
$^3${corresponding author, e-mail: stefano@duke.edu}
}
\date{\today}
\maketitle
\begin{abstract}
We report results of the first computer simulation studies of a
physically adsorbed gas on a quasicrystalline surface, Xe on
decagonal Al-Ni-Co.
The grand canonical Monte Carlo method is employed, 
using a semi-empirical gas-surface interaction, based on conventional combining rules, and 
the usual Lennard-Jones Xe-Xe interaction. The resulting adsorption isotherms and calculated 
structures are consistent with the results of {\small LEED} experimental data. 
The evolution of the bulk film begins in the second layer, while the low coverage behavior is epitaxial.  
This transition from 5-fold to 6-fold ordering is temperature dependent, 
occurring earlier (at lower coverage) for the higher temperatures. 
\end{abstract}
}

The observed unusual electronic\cite{ref1,ref2} and frictional\cite{ref3,ref4,ref5} properties of quasicrystal 
surfaces stimulate interesting fundamental questions about how these and other 
physical properties are altered by quasiperiodicity. Recent progress in the 
characterization and preparation of quasicrystal surfaces raises new possibilities 
for their use as substrates in the growth of films having novel structural, electronic, 
dynamic and mechanical properties\cite{ref6}, or as templates to facilitate the formation of 
well-ordered and/or size-selected nanoscale aperiodic arrays of clusters or molecules\cite{ref7,ref8}.

The physical behavior of systems involving competing interactions in adsorption is a 
subject of continuing interest\cite{ref9} and is particularly relevant to the growth of thin films. 
Several different growth modes have been observed for the growth of metal films on quasicrystals. 
For instance, Cu on Al-Pd-Mn grows layer-by-layer to produce a film having aperiodic order in one direction\cite{ref10}.  
Al on Al-Cu-Fe forms small size-selected nanoclusters at very low coverages\cite{ref11}. 
Ag on Al-Pd-Mn nucleates at strong-binding sites and ultimately grows as pyramidal hexagonal nanocrystals\cite{ref12}. 
Sb and Bi on Al-Pd-Mn and Al-Ni-Co form quasiperiodic monolayers\cite{ref13}. 
Some of these structures convert to different, often periodic, structures when annealed, 
and the issue of true equilibrium in the film is sometimes complicated by a tendency for 
chemical intermixing with the substrate. The wide range of behavior observed so far 
indicates that, even in the absence of intermixing, film growth is strongly affected by 
chemical interactions between adsorbate and substrate. In order to separate these chemical 
effects from those specific to quasiperiodic order, we have studied the adsorption of rare 
gases on a quasicrystal surface, where both the gas-gas and gas-surface interactions are 
believed to be simple, i.e., appreciable chemical interactions and adsorbate-induced 
surface reconstructions are absent.  

In the present work, we explore the implications of structural
mismatch by evaluating the nature of Xe adsorption on a quasicrystal
substrate, namely the 10-fold surface of decagonal Al-Ni-Co. 
This study of thermal and structural 
properties employs grand canonical Monte Carlo ({\small GCMC}) simulations, with which we have 
extensive experience\cite{ref14,ref15,ref16,ref17}. 
The calculations employ the same potential function we used 
earlier to compute the low coverage adsorption with the virial expansion\cite{ref18}. 
Using the {\small GCMC} method, we compute the film properties for specified thermodynamic conditions. 
A hard wall at 10 nm above the surface is used to confine the coexisting vapor phase. 
We take a square section of the surface $A$, of side 5.12 nm, to be the unit cell in the simulation, 
for which we assume periodic boundary conditions. This approach sacrifices accuracy of 
the long range QC structure. However, such a simplification is numerically useful for 
these simulations. 
Since the cell is large relative to the Xe size it is accurately representative of order on 
short-to-moderate length scales. The simulation results, presented below, are remarkably 
consistent with both the results from the virial calculations\cite{ref18} and with our experiment\cite{ref19}: 
a monolayer film is found, the ordering of which reflects many aspects of the underlying 
quasicrystalline structure. During the onset of second layer adsorption, there is a 
transition into a 6-fold structure, which after several layers is consistent with a bulk 
Xe structure.

\begin{figure}
 \centerline{\epsfig{file=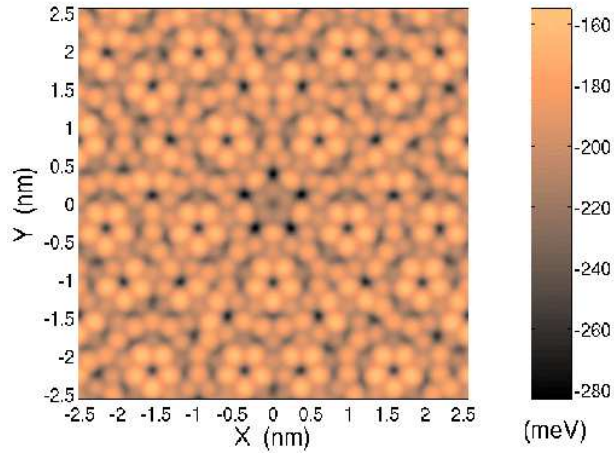,height=65mm,clip=}}
 \vspace{1mm}
 \caption{Computed potential energy for Xe on AlNiCo and potential energy scale (right), 
   obtained by minimizing $V(z,y,z)$ with respect to $z$ variation.}
 \label{fig1}
\end{figure}

The Xe-surface potential used here, shown in Figure \ref{fig1}, was calculated earlier by first 
developing an empirical adsorption potential, based on simple combining rules and known 
properties of bulk phases of solid Al, Ni and Co crystals\cite{ref18}. The potential is based on 
a summation of two-body interactions between the Xe and the individual constituent atoms 
of the substrate: Al, Ni and Co. The gas-gas potential is taken simply to be a 
Lennard-Jones (LJ) interaction, with parameter values for Xe: $\sigma=221$ K and $\epsilon=0.41$ nm. 
The Xe-substrate pair interactions are also assumed to have LJ form, with parameter 
values taken from traditional combining rules, using atomic sizes derived from bulk 
crystalline lattice constants\cite{ref18,ref19,ref20}. 
In the calculation of the adsorption potential, 
we assumed a structure of the virgin surface taken from an 
empirical fit to {\small LEED} data\cite{ref21}. 

\vspace{-5mm}
\vspace{-2mm}

The Xe adsorption potential derived with this procedure is both deep and highly corrugated. 
Depending on the lateral position $(x,y)$ across the surface, the maximum depth as a function 
of normal coordinate $(z)$ is typically in the range $D(x,y) \sim 150$ to $250$ meV. 
This potential is called ``deep'' because the record maximum well-depth for Xe on a periodic 
surface is about 160 meV, viz. on graphite\cite{ref22}; the record minimum well-depth 
is about 28 meV, on Cs\cite{ref23}. 
While the term ``corrugation'' is not well-defined for a non-periodic surface, one estimate 
of its magnitude comes from the standard deviation of the laterally varying well-depth, 
which is about 50 meV on this surface. Thus the well-depth's fluctuation is about 25\% of 
its average magnitude, which is sufficiently large as to warrant the description 
``highly corrugated'', in our opinion. Using this potential, we computed the adsorption 
properties at low vapor pressure, $P$, using the virial expansion (including the first three terms). 
The results of that analysis were found to be semi-quantitatively consistent with 
our experimental data in the low coverage regime\cite{ref18}.

\begin{figure}
 \centerline{\epsfig{file=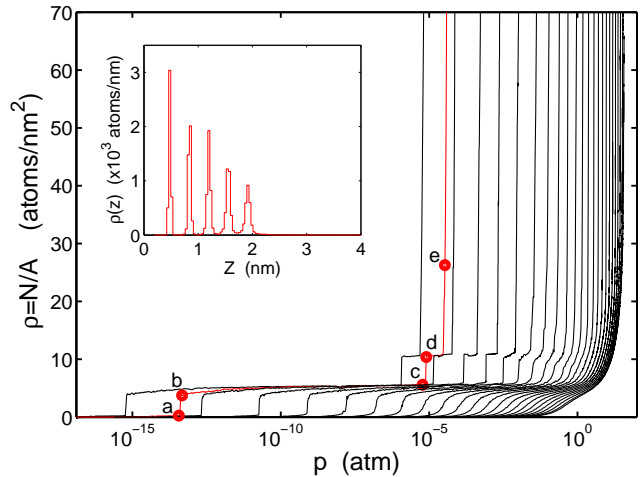,width=85mm,clip=}}
 \vspace{1mm}
 \caption{Computed isotherms from 70 K to 280 K in steps of 10 K.  
   Two additional isobars at 77 K and 286 K are shown.  
   The isotherm with solid circles is for 77 K, 
   while the highest-temperature isotherm is for 286 K.  
   The inset shows the density profile   
   $\rho(z)$ for $P=0.259 \times 10^{-4}$ atm and 77 K, 
   indicated by point (e).}
 \label{fig2}
\end{figure}

Figure \ref{fig2} shows adsorption isotherms computed with the {\small GCMC} simulations. 
The temperature range explored here extends from 70 K to 286 K 
(the triple temperature of Xe is 161.4 K). 
The plotted quantity is the thermodynamic excess coverage, $N_x/A$, 
defined as the difference between the total number of atoms in the simulation cell, 
and the number that would be present if the cell were filled with uniform vapor at the 
specified values of $P$ and $T$.  Detailed inspection of the isotherms reveals that there 
is continuous film growth (i.e. complete wetting) for all temperatures above the triple 
point. This behavior persists to some temperature below the triple temperature but we 
have not clearly established where the wetting transition occurs. At 77 K, at least 5 
layers form before the onset of bulk condensation. The inset in Figure \ref{fig2} shows the 
density profile, $\rho(z)$, in the direction perpendicular to the surface, for a total 
surface coverage of 26.28 atoms/nm$^2$, corresponding to point ``e'' on the 77 K isotherm. 
One observes that the (perpendicular) layer structure is well defined, 
in spite of the large lateral variation of the potential.  

\vspace{-3mm}
\vspace{-2mm}

\begin{figure}
 \centerline{\epsfig{file=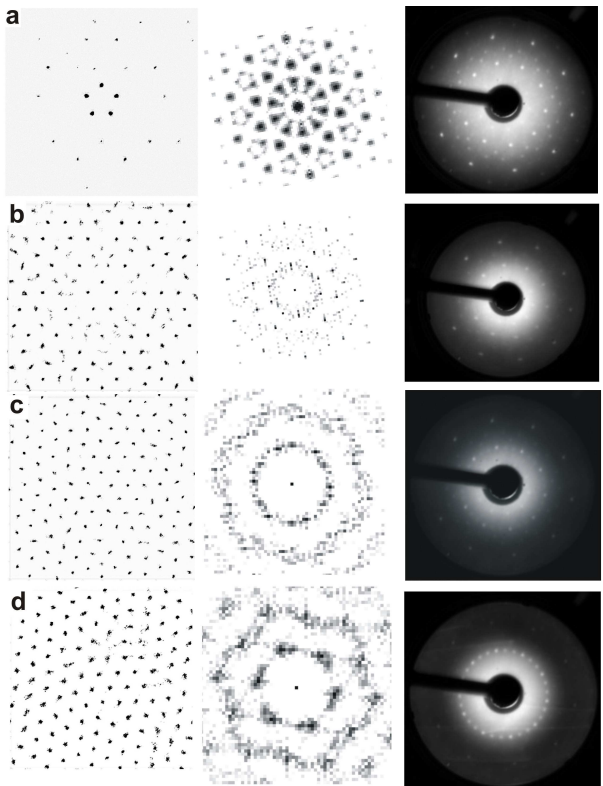,width=80mm,clip=}}
 \vspace{1mm}
 \caption{
   Intralayer density plots for adsorption at 77 K, 
   Fourier transforms of these plots and {\small LEED} patterns corresponding to similar conditions. 
   The plots show the density in the top layer, corresponding to the points noted in Figure \ref{fig2}.}
 \label{fig3}
\end{figure}

Figure \ref{fig3} shows the density variations within the top layer for several points on the 
77 K isotherm as specified in Figure \ref{fig2}. At the lowest pressure (point a), one observes 
a film that has atoms localized in the deepest parts of the potential, and the 
Fourier transform (FT) of that density function is 10-fold symmetric, reflecting the 
substrate symmetry. At the top of the first-layer step of the isotherm (point b), 
the density variation is more uniform, now having points of localized 5-fold and 6-fold 
symmetry. The FT of this structure is still 10-fold, i.e. still reflecting the substrate symmetry. 
By the time the adsorption proceeds to point c, however, the layer has 
more points of local 6-fold symmetry, and the FT displays 6-fold symmetry. 
When the full bilayer has formed (point d) the in-plane symmetry is clearly 6-fold 
locally, with some dislocation defects, and the FT indicates an overall 6-fold 
symmetry of the structure. When the 6-fold structure forms, it is aligned so that 
the close-packed direction of the Xe is parallel to a principal 5-fold direction of 
the substrate surface. For any given simulation run, the selection of 5-fold/6-fold 
alignment is arbitrary, and in Figure \ref{fig3}d, it can be seen that there are actually two alignments present.

Xe adsorption on this surface was studied earlier using low-energy electron diffraction, 
and isobar measurements indicate that the Xe film grows layer-by-layer in the 
temperature range 65K to 80K\cite{ref19}, consistent with the simulations described above. 
Figure \ref{fig3} shows the {\small LEED} patterns obtained from adsorbed Xe on decagonal Al-Ni-Co under 
similar conditions to the simulations. At the lowest coverage, the only discernible
change in the {\small LEED} pattern from that of the clean surface is an attenuation of the
substrate beams. After the adsorption of one layer (coverage determined by the isobar 
measurements\cite{ref19}) there are still no resolvable features that would indicate an overlayer 
having order different from the substrate. At the onset of the adsorption of the second 
layer, however, the {\small LEED} pattern shows new diffraction spots that correspond to 5 rotational 
domains of a hexagonal structure. Within each of these domains, the close-packed direction 
of the Xe is aligned with the 5-fold directions of the substrate\cite{ref19}, as observed in the 
simulation. In the experiments, all possible alignments are observed owing to the presence 
of all possible rotational alignments present within the width of the electron beam (0.25 mm), 
and there is no evidence of step-pinning of the overlayer. 
When the second layer is complete, these spots are well-defined and their widths are the 
same as the substrate spots, indicating a coherence length of at least 15 nm. 
A dynamical {\small LEED} analysis of the intensities indicates that the structure of the 
multilayer film is consistent with fcc Xe(111).

\vspace{-3mm}
\vspace{-2mm}

\begin{figure}
 \centerline{\epsfig{file=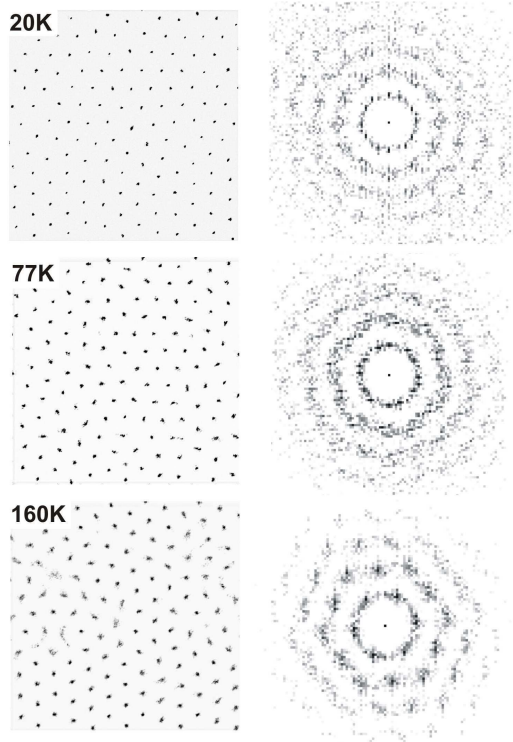,width=70mm,clip=}}
 \vspace{1mm}
 \caption{Density plots and Fourier transforms for monolayer films (corresponding to point c in Figure \ref{fig2}) at 20 K, 77 K, and 160 K.}
 \label{fig4}
\end{figure}

While the experiments are restricted to a comparatively narrow range in $T$ and $P$ 
(the {\small LEED} experiments require $P<10^{-7}$ bar and feasible equilibration times require $P>10^{-13}$ bar) 
the simulations are not so restricted. Therefore, we have also investigated the 
adsorption of Xe at a very low temperature (20 K) and a high temperature (160 K). 
The density plots and FT's at the coverage corresponding to point c in Figure \ref{fig3} is 
shown on Figure \ref{fig4} for the three temperatures, 20 K, 77 K and 160 K. The trend 
observed in the ordering is that the onset of 6-fold ordering occurs {\it earlier} at 
the higher temperatures. Six-fold ordering is already present at this coverage 
for 160 K, but is barely present at 77 K and is clearly not present at 20 K. 
This trend is consistent with the substrate potential being a bigger influence 
on the film's structure at lower $T$, implying that the development of the bulk Xe 
structure is entropy-driven; this is a kind of wetting transition of the bulk Xe 
phase, attributable to a diminishing interfacial free energy cost with increasing $T$. 
More thermal disorder is also evident in the film structures at higher $T$. 

\vspace{-3mm}
\vspace{-2mm}

Interestingly, at all $T$ studied, stacking faults are evident in the multilayer films. 
Their origin appears to be dislocations in the layers, which are most prevalent at 
the highest temperatures studied, as expected for entropic reasons. 
This is consistent with x-ray diffraction studies of the growth of Xe on Ag(111), 
where stacking faults were observed for Xe growth under various growth 
conditions\cite{ref24,ref25}, although the overall structure observed was fcc(111). 
Such a stacking fault is evident in Figure \ref{fig5}, which shows a superposition of
Xe layers 2 and 4 at point e in Figure \ref{fig2}. The coincidence of the atom locations 
in the top left part of this figure is consistent with an hcp structure 
(ABAB stacking) whereas the offset observed in the lower left is consistent 
with an fcc structure (ABC stacking).
We note that while bulk Xe has an fcc structure, and indeed an fcc structure 
was found for the multilayer film in the {\small LEED} study, calculations of the 
bulk structure using LJ pair potentials such as those employed here result 
in a more stable hcp structure \cite{ref26}.
This difference apparently arises from a neglect (using the pair potentials) 
of d-orbital overlap interactions, which are more effective in fcc than 
in hcp structures \cite{ref26}.
This model agrees quantitatively with measurements of energy differences 
in fcc and hcp Ar near its triple point \cite{ref27}. 
The presence of dislocations and disclinations in the monolayer films and 
their variation with temperature raises interesting questions concerning the 
possibility of dislocation-mediated melting {\cite{ref28,ref29,ref30}, 
but the limited simulation size precludes any such analysis. 
Larger simulations with Xe and different noble gases are intended for the future.

\vspace{-3mm}
\begin{figure}
 \centerline{\epsfig{file=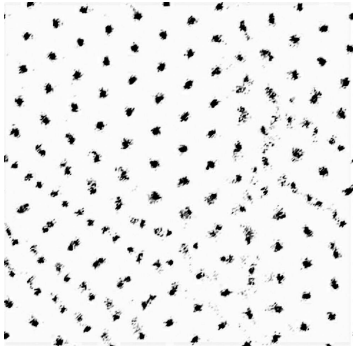,height=46mm,clip=}}
 \vspace{1mm}
 \caption{Density plot corresponding to point c on Figure \ref{fig2}, 
   showing a superposition of the density slices for the 2$^{nd}$ and 4$^{th}$ layers.  
   In the top right, 4$^{th}$-layer atoms are located directly 
   above the 2 $^{nd}$layer atoms, whereas in other regions, such as lower left, the two layers are offset.
}
 \label{fig5}
\end{figure}
\vspace{-2mm}

In summary, our simulation results are consistent with the experimental data 
over the somewhat limited range explored thus far. A consistent pattern has 
emerged; the quasicrystalline order is transmitted to films of thickness less 
than about two layers, disappearing upon further growth of the film. 
The results of this study encourage confidence in the use of physical 
adsorption to probe film growth. The extension of these calculations to 
the study of other properties such as lattice dynamics and friction, as 
well as other adsorbates, would seem well justified. 
Such studies will yield a comprehensive understanding of 
competing interactions in physisorbed layers on aperiodic substrates, 
as found previously on periodic surfaces.

This research was supported by NSF grant DMR-0208520. 
We wish to acknowledge helpful discussions with 
L. W. Bruch, Jorge Sofo, Mike Widom, Aleksey Kolmogorov, and R. Andreea Trasca.


\end{document}